\documentstyle[prl,twocolumn,aps,epsbox]{revtex}
\input{epsf.tex}

\begin{document} \draft 
\twocolumn[
\title{Mechanism of CDW-SDW Transition in One Dimension}
\author{Masaaki Nakamura\cite{email1}}
\address{Institute for Solid State Physics,University of Tokyo,
Roppongi, Tokyo 106-8666, Japan}
\date{\today}\maketitle
\begin{abstract}
\widetext\leftskip=0.10753\textwidth \rightskip\leftskip
 
 The phase transition between charge- and spin-density-wave (CDW, SDW)
 phases is studied in the one-dimensional extended Hubbard model at
 half-filling. We discuss whether the transition can be described by the
 Gaussian and the spin-gap transitions under charge-spin separation, or
 by a direct CDW-SDW transition. We determine these phase boundaries by
 level crossings of excitation spectra which are identified according to
 discrete symmetries of wave functions. We conclude that the Gaussian
 and the spin-gap transitions take place separately from weak- to
 intermediate-coupling region.  This means that the third phase exists
 between the CDW and the SDW states.  Our results are also consistent
 with those of the strong-coupling perturbative expansion and of the
 direct evaluation of order parameters.
 
\end{abstract}
] \narrowtext

%%% INTRODUCTION

 It has been pointed out that the phase transition between the
 charge-density-wave (CDW) and the spin-density-wave (SDW) phases of the 
 one-dimensional (1D) extended Hubbard model (EHM) has curious
 properties. The Hamiltonian of this model is given by
 \begin{eqnarray}
 {\cal H}&=&
 -t\sum_{is}(c^{\dag}_{is} c_{i+1,s}+\mbox{H.c.})\nonumber\\
 &&+U\sum_i n_{i\uparrow}n_{i\downarrow}+V\sum_i n_{i}n_{i+1}.
 \label{eqn:tUV}
 \end{eqnarray}
 From the strong coupling theory, it has been shown that the first-order
 transition takes place near $U=2V$\cite{Bari,Hirsch,Dongen}. On the
 other hand, the weak coupling theory predicts that the transition is
 the second order on the same line\cite{Emery,Solyom}. This means that
 there exist a crossover between these two transitions while, in more
 than two dimension, the transition is the first order for any strength
 of the interactions.  Many authors investigated this transition by
 analytical\cite{Voit92} and numerical
 approaches\cite{Hirsch,Fourcade-S,Cannon-F,Cannon-S-F,GPZhang}.
 Numerical analysis has been done by the direct estimation of the CDW
 order parameter\cite{Hirsch,Cannon-F,Cannon-S-F} and by the real-space
 renormalization group technique\cite{Fourcade-S}. However, due to the
 lack of precision of the analysis, the property of this transition
 is still left to be ambiguous.
 
 In this Letter, we study this transition by investigating
 level-crossings of excitation spectra
 \cite{Julien-H,Okamoto-N,Nomura-O,Nakamura-N-K,Nakamura}
 which enable us to determine the phase boundary with high accuracy from
 the numerical data of finite-size clusters. Using this technique, we
 determine the phase boundaries as shown in Fig.\ref{fig:3lines}. Then
 we conclude that two transitions of charge and spin degrees of freedom
 occur independently, and the third phase exists between the CDW and the
 SDW phases.

%%% METHODS

 First, we briefly explain the level-crossing method used in our
 analysis. In general, the low-energy behavior of 1D electron systems
 can be described as Tomonaga-Luttinger (TL) liquids \cite{Voit}. In
 the TL liquid (bosonization) theory, the continuous fermion fields are
 defined by $c_{is}\rightarrow \psi_{{\rm L},s}(x)+\psi_{{\rm R},s}(x)$
 with
 \begin{equation}
  \psi_{r,s}(x)=\frac{1}{\sqrt{2\pi\alpha}}
   {\rm e}^{{\rm i} r k_{\rm F} x}{\rm e}^{{\rm i}/\sqrt{2}\cdot
  [r(\phi_{\rho}+s\phi_{\sigma})-\theta_{\rho}-s\theta_{\sigma}]},
   \label{eqn:fermion_op}
 \end{equation}
 where $r={\rm R},{\rm L}$ and $s=\uparrow, \downarrow$ refer to $+,-$
 in that order. $k_{\rm F}$ is the Fermi wave number. The field
 $\phi_{\nu}$ and dual field $\theta_{\nu}$ of the charge ($\nu=\rho$)
 and spin ($\nu=\sigma$) degrees of freedoms satisfy the relation
 $[\phi_{\mu}(x),\theta_{\nu}(x')]=-{\rm i}\pi\,\delta_{\mu\nu}{\rm
 sign}(x-x')/2$. Then the effective Hamiltonian for the system with
 length $L$ is given by the sine-Gordon model:
 \begin{eqnarray}
   \lefteqn{
    {\cal H}=\sum_{\nu=\rho,\sigma}\frac{v_{\nu}}{2\pi}\int_0^L {\rm d}x
  \left[K_{\nu}(\partial_x \theta_{\nu})^2
       +K_{\nu}^{-1}(\partial_x \phi_{\nu})^2\right]}\label{eqn:eff_Ham}\\
    &&+\frac{2 g_{3\perp}}{(2\pi\alpha)^2}
  \int_0^L {\rm d}x \cos[\sqrt{8}\phi_{\rho}]
     +\frac{2 g_{1\perp}}{(2\pi\alpha)^2}
  \int_0^L {\rm d}x \cos[\sqrt{8}\phi_{\sigma}],\nonumber
 \end{eqnarray}  
 where $v_{\nu}$ and $K_{\nu}$ denote the sound velocity and the
 Gaussian coupling, respectively, for each sector. The non-linear terms
 in eq.(\ref{eqn:eff_Ham}) are originated from the Umklapp (charge) and
 the backward (spin) scattering effects which can cause the charge- and
 the spin-gap instabilities. If these non-linear terms are irrelevant,
 the excitation spectra and their wave numbers in this system are given
 by the relation
 \begin{eqnarray}
  \Delta E&=&
  \frac{2\pi v_{\rho}}{L}x_{\rho}+\frac{2\pi v_{\sigma}}{L}x_{\sigma},\\
  k&=&\frac{2\pi}{L}(s_{\rho}+s_{\sigma})+2m_{\rho}k_{\rm F},
 \end{eqnarray} 
 where $x_{\nu}=(n_{\nu}^2/K_{\nu}+m_{\nu}^2K_{\nu})/2$ are the scaling
 dimensions and $s_{\nu}=n_{\nu}m_{\nu}$ are the conformal
 spins. The integers $n_{\nu}$ and $m_{\nu}$ are quantum numbers for the
 particle numbers and the current excitations, respectively. The scaling
 dimensions are related to the critical exponents for the correlation
 functions as
 \begin{equation}
  \langle{\cal O}_i(r){\cal O}_i(r')\rangle
  \sim |r-r'|^{-2(x_{\rho i}+x_{\sigma i})}.
 \end{equation}
 Therefore, there is one to one correspondence between the excitation
 spectra and the operators.

 In our analysis we turn our attention on excitation spectra which
 correspond to the following operators:
 \begin{mathletters}
 \begin{eqnarray}
 {\cal O}_{\nu 1}&\equiv&\sqrt{2}\cos(\sqrt{2}\phi_{\nu}),\label{eqn:cos}\\
 {\cal O}_{\nu 2}&\equiv&\sqrt{2}\sin(\sqrt{2}\phi_{\nu}),\label{eqn:sin}\\
 {\cal O}_{\nu 3}&\equiv&\exp(\pm{\rm i}\sqrt{2}\theta_{\nu}).\label{eqn:exp}
 \end{eqnarray}\label{eqn:ops}
 \end{mathletters}
 For the spin sector ($\nu=\sigma$) which have an SU(2) symmetry, the
 operators (\ref{eqn:cos}) and (\ref{eqn:sin},\ref{eqn:exp}) form
 singlet and triplet states, respectively.  In this case, the level
 crossing of these spectra ($x_{\sigma1}=x_{\sigma2},x_{\sigma3}$) gives 
 the spin-gap phase boundary\cite{Okamoto-N,Nakamura-N-K}. On the other
 hand, the charge sector ($\nu=\rho$) is U(1) symmetric. Then the
 operators of eqs.(\ref{eqn:ops}) correspond to ``N\'{e}el'', ``dimer'',
 and ``doublet'' states in that order, borrowing the terminology of
 anisotropic spin chains.  In this case the level-crossing of
 ``N\'{e}el'' and ``dimer'' excitations ($x_{\rho 1}=x_{\rho 2}$) gives
 the Gaussian transition\cite{Nomura-O,BKT}, which means a second order
 transition between two massive states with different fixed points
 ($g_{3\perp}\rightarrow\pm\infty$). Note that these excitation spectra
 can be extracted when we choose anti-periodic boundary conditions
 (${\rm BC}=-1$, see Tab. \ref{tbl:symmetries}), reflecting the
 selection rule of the quantum numbers\cite{Nakamura-N-K}. In the
 weak-coupling limit, the Gaussian and the spin-gap transition take
 place simultaneously, because $g_{1\perp}=g_{3\perp}=U-2V$. However,
 there is no guarantee for the synchronization except for this limit.

%% Charge-Spin Coupling

 In addition to the above effective Hamiltonian (\ref{eqn:eff_Ham}),
 there exists the following Umklapp operator transferring finite spin
 \cite{Voit92,Cannon-F,Kolomeisky-S}:
 \begin{equation}
  {\cal H}'=\frac{2 g_{3\parallel}}{(2\pi \alpha)^2}
  \int_0^L {\rm d}x
  \cos[\sqrt{8}\phi_{\rho}]\cos[\sqrt{8}\phi_{\sigma}].
  \label{eqn:Umklapp_spin}
 \end{equation}
 In the weak coupling limit, the coupling constant is assigned as
 $g_{3\parallel}=-2V$, so that it remains finite on the $U=2V$ line.
 Therefore, we should consider the possibility that the charge and the spin
 degrees of freedom is not separated, and a direct transition between
 the CDW and the SDW phases takes place. To examine this possibility, we
 also observe the level-crossing of excitation spectra of the CDW and
 the SDW operators which consist of both charge and spin components (see
 Tab. \ref{tbl:symmetries}). These excitation spectra can be obtained in
 periodic boundary conditions (${\rm BC}=1$) with the wave number
 $2k_{\rm F}=\pi$.

%% Discrete Symmetries

 The excitation spectra correspond to the above operators can be
 identified according to their discrete symmetries. The wave functions
 for the excited states can change their signs under particle-hole
 (${\cal C}$: $c_{is}\leftrightarrow(-1)^i c_{is}^{\dag}$),
 space-inversion (${\cal P}$: R$\leftrightarrow$L), and spin-reversal
 (${\cal T}$: $\uparrow\leftrightarrow\downarrow$) transformations. It
 follows from eq.(\ref{eqn:fermion_op}), that the phase fields
 $\phi_{\nu}$ change by these transformations as follows:
 \begin{mathletters}
 \begin{eqnarray}
  {\cal C}, {\cal P}:&&\phi_{\sigma}\rightarrow-\phi_{\sigma},\ \ \
            \phi_{\rho}\rightarrow\pi/\sqrt{2}-\phi_{\rho}
	    \label{eqn:phase_CP}\\
  {\cal T}:&&\phi_{\sigma}\rightarrow-\phi_{\sigma}.
 \end{eqnarray}
 \end{mathletters}
 Here, ${\cal C}{\cal P}=1$ is always satisfied, so that independent
 discrete symmetries are ${\cal P}$ and ${\cal T}$.  The boson
 representation of the operators and their symmetries are summarized in
 Tab. \ref{tbl:symmetries}\cite{operators}.  In the present numerical
 calculation based on the Lanczos algorism, the identification is
 done by projecting the initial vector as
 \begin{equation}
  |\Psi_{\rm init}\rangle=\frac{1}{2}(1\pm{\cal P})(1\pm{\cal T})|i\rangle,
 \end{equation}
 where the signs in front of the operators correspond to their eigen
 values, and $|i\rangle$ is some configuration which satisfies ${\cal
 P}, {\cal T}|i\rangle\neq\pm|i\rangle$. Furthermore, $|i\rangle$ is
 classified by the wave numbers $k=0,\pi$.
  
%%% RESULT

%% Finite-Size Effect and Consistency Check

\begin{figure}[t]
%* mono
%\epsfxsize=3.5in \leavevmode \epsfbox{diff.eps}
%* color
\epsfxsize=3.5in \leavevmode \epsfbox{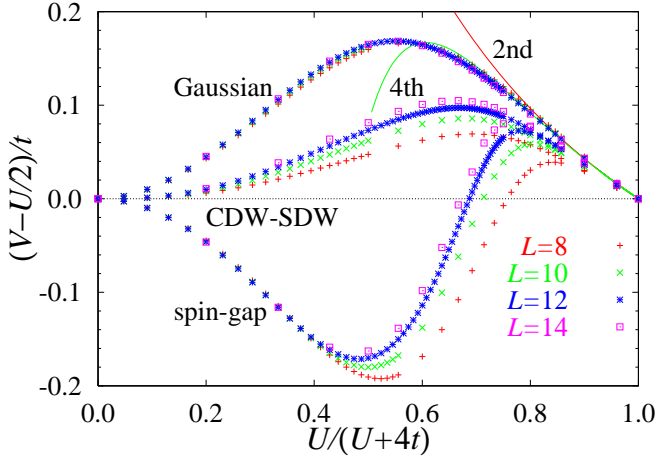}
\caption{Possible three transitions (Gaussian, spin-gap, and CDW-SDW
 transitions) along $U=2V$ line of the EHM calculated in $L=8,10,12,14$
 systems. The result of the strong coupling expansion agrees with the
 Gaussian transition. This means that the actual transitions are the
 Gaussian and the spin-gap transitions, and a BOW state exists between
 them.}
\label{fig:3lines}
\end{figure}

\begin{figure}[h]
\epsfxsize=3.5in \leavevmode \epsfbox{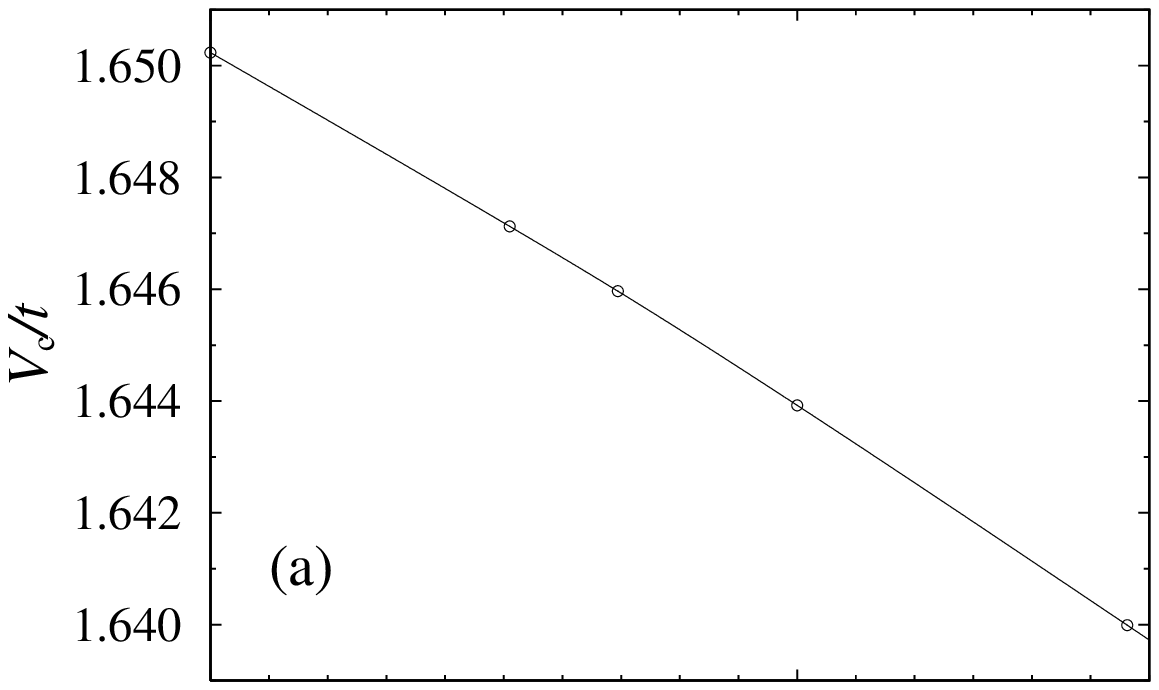}\\
\epsfxsize=3.5in \leavevmode \epsfbox{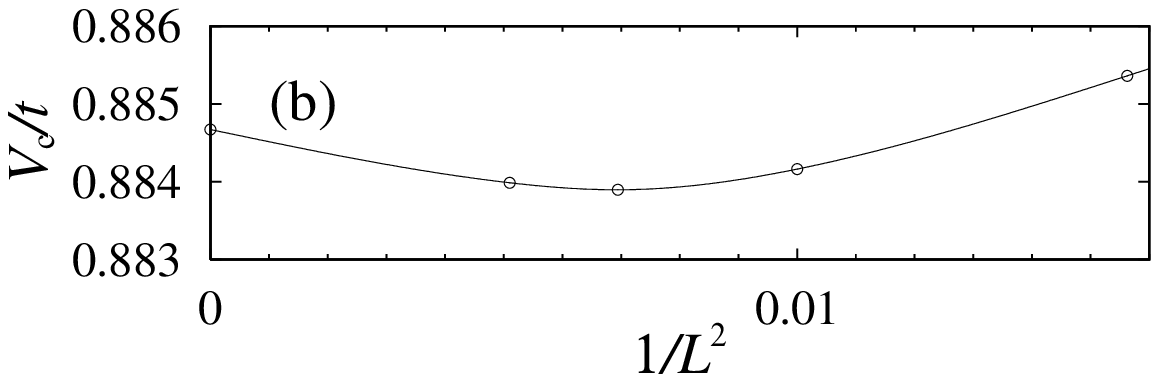}
\caption{Size dependence of the critical points for the (a) Gaussian
 transition at $U/t=3$, and the (b) spin-gap transition at $U/t=2$. The
 former agrees with the Cannon {\it et al.}'s result that
 $V_{\rm c}/t=1.65^{+0.10}_{-0.05}$\protect{\cite{Cannon-S-F}}.}
 \label{fig:size}
\end{figure}

 The critical lines obtained by the above explained way with the exact
 diagonalization of the $L=8,10,12,14$ systems are shown in
 Fig. \ref{fig:3lines}. For the Gaussian transition line, the
 finite-size effect is small for all region (see
 Fig.\ref{fig:size}(a)). On the other hand, for the spin-gap transition
 line, the finite-size effect is small in the weak-coupling region (see
 Fig.\ref{fig:size}(b)), but large in the intermediate- and the
 strong-coupling regime.  The CDW-SDW transition line lies between the
 above two lines\cite{DMRG}. Its finite-size effect is large for all
 region.

\begin{figure}[h]
\epsfxsize=3.5in \leavevmode \epsfbox{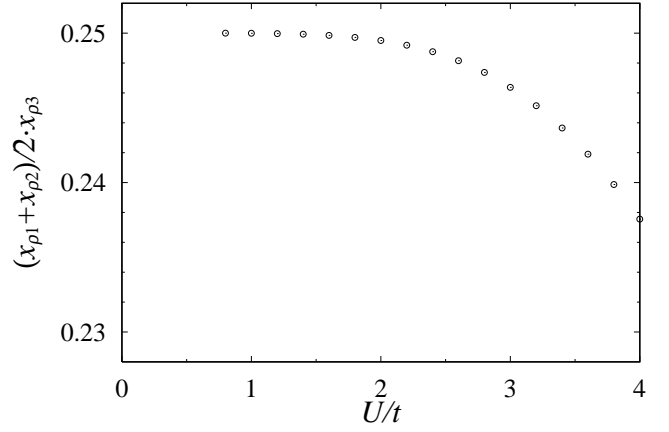}
\caption{Product of the scaling dimensions on the Gaussian critical line
 (eq.(\protect{\ref{eqn:check_ga}})). The TL liquid theory predicts the
 value takes $1/4$.}
\label{fig:check_ga}
\end{figure}

\begin{figure}[h]
\epsfxsize=3.5in \leavevmode \epsfbox{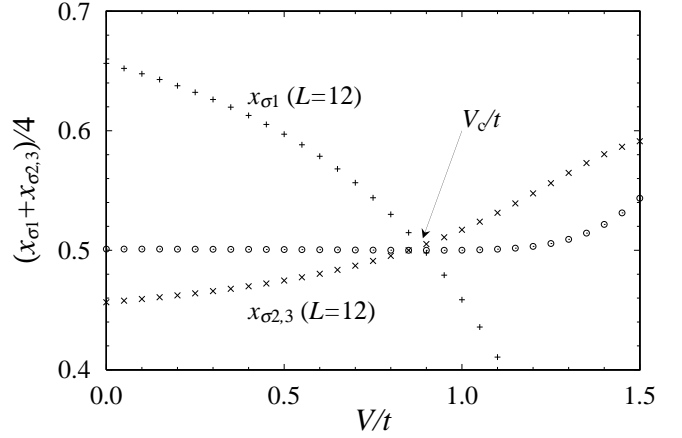}
\caption{Averaged scaling dimension of the spin sector near the spin-gap
 critical point at $U/t=2$ (eq.(\protect{\ref{eqn:check_sg}})). The TL
 liquid theory predicts the value takes $1/2$.}
\label{fig:check_sg}
\end{figure}

 In order to check the consistency of our argument, we confirm the
 relations between the scaling dimensions for each instability.
 The relation on the Gaussian critical line\cite{Nomura-O} and that near
 the spin-gap transition\cite{Okamoto-N,Nakamura-N-K} are given by
\begin{eqnarray}
  \frac{x_{\rho 1}+x_{\rho 2}}{2}x_{\rho 3}&=&\frac{1}{4},
   \label{eqn:check_ga}\\
  \frac{x_{\sigma 1}+3x_{\sigma 2,3}}{4}&=&\frac{1}{2}.
   \label{eqn:check_sg}
\end{eqnarray}
 The numerical results are shown in Figs.\ref{fig:check_ga} and
 \ref{fig:check_sg}. The Gaussian and the spin-gap transition lines
 satisfy the consistency of our theoretical scheme from the weak- to the
 intermediate coupling region.

%% Strong Coupling Theory

 Moreover, in order to back up the present result from the strong
 coupling theory, we compare above result to the critical line obtained
 by the strong coupling expansions. In this limit, the phase boundary
 between the CDW and the SDW states can be determined by equating
 energies of these states. This calculation has already been done by
 Hirsch\cite{Hirsch} and van Dongen\cite{Dongen} up to second and forth
 order, respectively, using the Bethe-ansatz results. Among the three
 transition lines we assumed, van Dongen's result shows good agreement
 with the Gaussian transition in the charge part up to $U/t\sim 6$.
 We should also note that our Gaussian critical point agrees with the
 Cannon {\it et al.}'s result obtained by the direct evaluation of the
 CDW order parameter: $V_{\rm c}/t=1.65^{+0.10}_{-0.05}$ for $U/t=3$
 $V_{\rm c}/t=2.92\pm 0.04$ for $U/t=5.5$\cite{Cannon-S-F} (see
 Fig. \ref{fig:3lines} and \ref{fig:size}(a)).

%%% CONCLUSION and DUSCUSSION
 
%% Crossover

 From the above evidence, we conclude that the actual transition near
 $U=2V$ line is not direct transition between the CDW and the SDW
 states, but independent Gaussian and spin-gap transitions at least from
 the weak- to the intermediate-coupling region. In the strong coupling
 regime, these two boundaries approach and coincide at the finite
 strength of the coupling. Unfortunately, in the present analysis, we
 can not determine this point, but it is considered to be identical to
 the crossover point between the second and the first order
 transitions. In this way, our analysis suggests that the crossover
 along the $U=2V$ line is closely related to the validity of the
 charge-spin separation.

%% BOW state

 Our result also means that there is a finite region with charge- and
 spin-gapped state which has different symmetries from the CDW
 state. This third phase is considered as a bond-order-wave (BOW) state
 with LRO which is characterized by the following operator:
\begin{equation}
 {\cal O}_{\rm BOW}=\frac{(-1)^i}{2}
 \sum_s(c^{\dag}_{i+1,s}c_{is}+c^{\dag}_{is}c_{i+1,s}).
\end{equation}
 Therefore the direct evaluation of this order parameter may possible.
 The existence of the BOW state can be more clarified by extending
 the EHM. According to the bosonization analysis of the EHM with
 correlated-hopping interactions\cite{Japaridze-K}, a finite BOW region
 remains even in the weak coupling limit ($g_{1\perp} < g_{3\perp}$). On
 the other hand, if the Gaussian and the spin-gap transitions take place
 in the opposite order ($g_{1\perp} > g_{3\perp}$), there appears a
 bond-spin-density-wave (BSDW) phase which has massive charge sector and 
 massless spin sector.
   
% Other systems

 Finally, we refer to other examples of the crossover similar to that in
 the EHM. This type of phenomenon has also been observed in the
 transition between the singlet and the Haldane phases in the $S=1/2$
 frustrated spin-ladder model\cite{XWang}. Our result would also shed
 light on such a transition.

%%% SUMMARY

 In summary, we have studied the CDW-SDW transition in the EHM along the
 $U=2V$ line, and shown that there exist the Gaussian and the spin-gap
 transitions in the charge and the spin degrees of freedom,
 respectively, and there is the BOW state between them. The crossover
 from the second to the first order transition is suggested to be
 related with the validity of the charge-spin separation.

%%% ACKNOWLEDGMENT

 The author is grateful to T. Kawarabayashi and H. Otsuka for
 stimulating discussions. He also thanks K. Itoh for drawing his
 attention to the correlated hopping model. The computation in this work 
 was partly done using the facilities of the Supercomputer Center,
 Institute for Solid State Physics, University of Tokyo.

\pagebreak
\begin{table}[h]
\begin{tabular}{c|c|rrrrr}
 &operators&${\cal C}$& ${\cal P}$& ${\cal T}$&  $k$& BC\\ \hline
 G.S.
 & $1$
 & $ 1$& $ 1$ & $ 1$ & $0$ & $\pm 1$\\
 marginal
 & $-\frac{4}{K_{\nu}}\bar{\partial}\phi_{\nu}\partial\phi_{\nu}$
 & $ 1$& $ 1$ & $ 1$ & $0$ & $\pm 1$\\
 BOW
 & $\sin\sqrt{2}\phi_{\rho}\cdot\cos\sqrt{2}\phi_{\sigma}$
 & $ 1$& $ 1$ & $ 1$ & $2k_{\rm F}$ & $\pm 1$\\
 SDW$_{zz}$
 & $\sin\sqrt{2}\phi_{\rho}\cdot\sin\sqrt{2}\phi_{\sigma}$
 & $-1$& $-1$ & $-1$ & $2k_{\rm F}$ & $\pm 1$\\
 SDW$_{xx}$
 & $\sin\sqrt{2}\phi_{\rho}\cdot\exp\pm{\rm i}\sqrt{2}\theta_{\sigma}$
 & $ *$& $ 1$ & $ *$ & $2k_{\rm F}$ & $\pm 1$\\
 CDW
 & $\cos\sqrt{2}\phi_{\rho}\cdot\cos\sqrt{2}\phi_{\sigma}$
 & $-1$& $-1$ & $ 1$ & $2k_{\rm F}$ & $\pm 1$\\
 BSDW$_{zz}$
 & $\cos\sqrt{2}\phi_{\rho}\cdot\sin\sqrt{2}\phi_{\sigma}$
 & $ 1$& $ 1$ & $-1$ & $2k_{\rm F}$ & $\pm 1$\\
 BSDW$_{xx}$
 & $\cos\sqrt{2}\phi_{\rho}\cdot\exp\pm{\rm i}\sqrt{2}\theta_{\sigma}$ 
 & $ *$& $-1$ & $ *$ & $2k_{\rm F}$ & $\pm 1$\\
 SS
 & $\exp{\rm i}\sqrt{2}\theta_{\rho}\cdot\cos\sqrt{2}\phi_{\sigma}$
 & $ *$& $ 1$ & $ 1$ & $0$ & $\pm 1$\\
 TS$_0$
 & $\exp{\rm i}\sqrt{2}\theta_{\rho}\cdot\sin\sqrt{2}\phi_{\sigma}$
 & $ *$& $-1$ & $-1$ & $0$ & $\pm 1$\\
 TS$_1$
 &$\exp{\rm i}\sqrt{2}\theta_{\rho}\cdot\exp\pm{\rm i}\sqrt{2}\theta_{\sigma}$
 & $ *$& $ 1$ & $ *$ & $0$ & $\pm 1$\\
 $4k_{\rm F}$-CDW
 & $\cos 2\sqrt{2}\phi_{\rho}$
 & $ *$& $-1$ & $ *$ & $4k_{\rm F}$ & $\pm 1$\\
\hline
 singlet
 & $\cos\sqrt{2}\phi_{\sigma}$
 & $ 1$& $ 1$ & $ 1$ & $0$ & $\mp 1$\\
 triplet$_0$
 & $\sin\sqrt{2}\phi_{\sigma}$
 & $-1$& $-1$ & $-1$ & $0$ & $\mp 1$\\
 triplet$_1$
 & $\exp\pm{\rm i}\sqrt{2}\theta_{\sigma}$
 & $ *$& $ 1$ & $ *$ & $0$ & $\mp 1$\\
 ``dimer'' &
 $\sin\sqrt{2}\phi_{\rho}$
 & $ 1$& $ 1$ & $ *$ & $2k_{\rm F}$ & $\mp 1$\\
 ``N\'{e}el''
 & $\cos\sqrt{2}\phi_{\rho}$
 & $-1$& $-1$ & $ *$ & $2k_{\rm F}$ & $\mp 1$\\
 ``doublet''
 & $\exp\pm{\rm i}\sqrt{2}\theta_{\rho}$
 & $ *$& $ 1$ & $ 1$ & $0$ & $\mp 1$
\end{tabular}
\caption{Discrete symmetries of wave functions which correspond to
 several excitation spectra (${\cal C}$: charge conjugation, ${\cal P}$:
 space inversion, ${\cal T}$: spin reversal, and $k$: wave number). The
 upper (lower) sign of boundary conditions (BC) denotes $N/2=$odd (even)
 cases. The upper 12 states are ``physical'' states which appear in the
 same boundary conditions as the ground state. The lower 6 states are
 the ``artificial'' ones which are extracted by twisting the boundary
 conditions respect to the ground state.}
\label{tbl:symmetries}
\end{table}

\begin{references}
\bibitem{DMRG}
 In Ref.10, the critical point is determined by observing
 the CDW and the SDW structure factor using the
 density-matrix-renomalization-group method. This result agrees with
 our level-crossing point of the CDW and the SDW excitations.
\end{references}
\end{document}